\documentclass[10pt,twocolumn,floatfix,superscriptaddress,preprintnumbers,amsmath,amssymb,prb]{revtex4}
\usepackage{graphicx,bm}
\usepackage{amssymb,amsmath}
\usepackage{float}
\usepackage{epstopdf}
\usepackage[dvips]{color}

\newcommand*{\citen}[1]{%
  \begingroup
    \romannumeral-`\x 
    \setcitestyle{numbers}%
    \cite{#1}%
  \endgroup   
}

\begin{document}

\title{Superconducting and normal-state properties of the noncentrosymmetric superconductor Re$_{6}$Zr}

\author{D. A. Mayoh}
\email[]{d.mayoh@warwick.ac.uk}
\affiliation{Physics Department, University of Warwick, Coventry, CV4 7AL, United Kingdom}

\author{J. A. T. Barker}
\altaffiliation[]{Present address: Paul Scherrer Institut, 5232 Villgen PSI, Switzerland}
\affiliation{Physics Department, University of Warwick, Coventry, CV4 7AL, United Kingdom}

\author{R. P. Singh}
\affiliation{Department of Physics, Indian Institute of Science Education and Research Bhopal, Bhopal 462066, India}

\author{G. Balakrishnan}
\affiliation{Physics Department, University of Warwick, Coventry, CV4 7AL, United Kingdom}

\author{D. McK. Paul}
\affiliation{Physics Department, University of Warwick, Coventry, CV4 7AL, United Kingdom}

\author{M. R. Lees}
\email[]{m.r.lees@warwick.ac.uk}
\affiliation{Physics Department, University of Warwick, Coventry, CV4 7AL, United Kingdom}

\begin{abstract}
We systematically investigate the normal and superconducting properties of non-centrosymmetric Re$_{6}$Zr using magnetization, heat capacity, and electrical resistivity measurements. Resistivity measurements indicate Re$_{6}$Zr has poor metallic behavior and is dominated by disorder. Re$_6$Zr undergoes a superconducting transition at $T_{\mathrm{c}} = \left(6.75\pm0.05\right)$~K. Magnetization measurements give a lower critical field, $\mu_{0}H_{\mathrm{c1}} = \left(10.3 \pm 0.1\right)$~mT. The Werthamer-Helfand-Hohenberg model is used to approximate the upper critical field $\mu_{0}H_{\mathrm{c2}} = \left(11.2 \pm 0.2\right)$~T which is close to the Pauli limiting field of 12.35 T and which could indicate singlet-triplet mixing. However, low-temperature specific-heat data suggest that Re$_{6}$Zr is an isotropic, fully gapped $s$-wave superconductor with enhanced electron-phonon coupling. Unusual flux pinning resulting in a peak effect is observed in the magnetization data, indicating an unconventional vortex state.
\end{abstract}

\maketitle
\section{INTRODUCTION}

In superconductors, the inversion symmetry of the crystallographic structure plays a central role in the formation of the Cooper pairs. In conventional superconductors, each Cooper pair is formed from two electrons which belong to the same Fermi surface with a symmetric orbital state and an antisymmetric spin state. The discovery of superconductivity in CePt$_3$Si, a material which lacks inversion symmetry, has generated considerable experimental and theoretical interest in the physics of noncentrosymmetric (NCS) superconductors~\cite{Bauer04,Bauer07,Bauer12}. The absence of inversion symmetry in NCS materials introduces an antisymmetric spin-orbit coupling~\cite{Rashba,Rashba1} which can result in a splitting of the spin-up and spin-down conduction electron energy bands. This splitting of the Fermi surface, lifting the degeneracy of the conduction electrons, may result in a superconducting pair wave function that is an admixture of spin-singlet and spin-triplet states, although there are several examples of NCS superconductors where it has been established that the order parameter is not unconventional, for example, BiPd~\cite{Sun} and PbTaSe$_{2}$~\cite{Guan}.  Singlet-triplet mixing can lead NCS materials to display significantly different properties from conventional superconducting systems, for example, the triplet pairing seen in Li$_2$(Pd,Pt)$_3$Si~\cite{Yuan2006, Nishiyama2007, Takeya2007, Harada2012}, and upper critical fields close to or exceeding the Pauli limiting field observed in Mo$_{3}$Al$_{2}$C ~\cite{Mo3Al2C}~\cite{Bauer12} Re$_{3}$W~\cite{ReW}, Ca(Ir,Pt)Si$_{3}$~\cite{CaIr}, Li$_2$(Pd,Pt)$_3$Si~\cite{Yuan2006, Nishiyama2007, Takeya2007, Harada2012}, LaRhSi$_{3}$~\cite{LaRhSi3}, Nb$_{0.18}$Re$_{0.82}$~\cite{NbRe}, Y$_{2}$C$_{3}$~\cite{Chen}, and Mg$_{10}$Ir$_{19}$B$_{16}$~\cite{Mg10Ir19B16}.



Noncentrosymmetric superconductors are prime candidates to exhibit time-reversal symmetry (TRS) breaking. Until recently, however, this rare phenomenon had been observed directly in only a few unconventional centrosymmetric superconductors, for example, PrPt$_{4}$Ge$_{12}$~\cite{PrPt4Ge12}, Sr$_{2}$RuO$_{4}$~\cite{Sr2RuO4,OpticalSr2RuO4}, (Pr,La)(Os,Ru)$_{4}$Sb$_{12}$~\cite{PrOsSb,PrOsRuSb}, UPt$_{3}$ and (U,Th)Be$_{13}$~\cite{UPt3,OneUPt3,TwoUPt3,UThBe}, and LaNiGa$_{2}$~\cite{LaNiGa2} and the cage-type superconductors Lu$_5$Rh$_6$Sn$_{18}$~\cite{Lu5Rh6Sn18}, while no spontaneous magnetization associated with TRS breaking had been reported in any of the NCS materials mentioned above. 

Recently, this situation has changed, and several NCS superconductors have been reported to show TRS breaking. In the first of these, LaNiC$_{2}$, symmetry analysis implies that the superconducting instability is of the nonunitary triplet type, with a spin-orbit coupling that is comparatively weak and with mixing of singlet and triplet pairing being forbidden by symmetry~\cite{LaNiC2, LaNiC2Theo}. TRS breaking was also found in La$_7$Ir$_3$, with measurements of the superconducting gap indicating that it is isotropic with a superconducting ground state that is dominated by an $s$-wave component~\cite{La7Ir3}.


Re$_{6}$Zr is a member of the $\alpha$-Mn family of intermetallic compounds~\cite{Matthias} and has a noncentrosymmetric cubic structure, space group $I\bar{4}3m$. We have previously reported the results of muon spin relaxation ($\mu$SR) measurements on Re$_{6}$Zr, showing that TRS is broken in this material. A theoretical analysis of the possible pairing states demonstrated that a mixing of spin-singlet and spin-triplet pairing is possible in this noncentrosymmetric superconducting compound~\cite{Re6Zr,Matthias}. Here, we present a comprehensive characterization of the normal and superconducting states of this intermetallic compound through studies by magnetization, electronic transport, and heat capacity. We estimate several normal state parameters of Re$_{6}$Zr such as the electronic specific-heat contribution $\gamma_{\mathrm{n}}$, residual resistivity $\rho_{\mathrm{0}}$, and the hyperfine contribution to the specific heat. Using the electronic-transport and heat-capacity measurements, we estimate the Debye temperature by using the parallel-resistor model, the Debye lattice contribution to the specific heat at low temperature, and the Debye-Einstein model. Several superconducting parameters, including the lower and upper critical fields $H_{\mathrm{c}1}$ and $H_{\mathrm{c}2}$, the coherence length $\xi_{\mathrm{GL}}$, and the penetration depth $\lambda_{\mathrm{GL}}$, are estimated. The specific-heat jump $\Delta C/\gamma_{\mathrm{n}} T_{\mathrm{c}}$, the superconducting gap $\Delta_{0}/k_{\mathrm{B}}T_{\mathrm{c}}$, and the temperature dependence of the specific heat at low-temperature suggest that Re$_{6}$Zr is an isotropic, fully gapped $s$-wave superconductor with enhanced electron-phonon coupling. We also present evidence of unusual flux pinning not normally seen in low-$T_{\mathrm{c}}$ systems.



\section{Experimental Details}

Polycrystalline samples of Re$_{6}$Zr were prepared by arc melting stoichiometric quantities of high-purity ($4N$) Zr and Re in an arc furnace under an argon ($5N$) atmosphere on a water-cooled copper hearth. The sample buttons were melted and flipped several times to ensure phase homogeneity. The observed weight loss during the melting was negligible. Powder x-ray diffraction data confirmed the $\alpha$-Mn crystal structure and the phase purity of the samples. A low ($\chi_{\mathrm{dc}}=5.8\times 10^{-4}$), nearly temperature independent normal-state dc susceptibility indicates there are no magnetic impurities from the Zr.

The normal and superconducting states of Re$_{6}$Zr were characterized by magnetization $M$, ac susceptibility $\chi_{\mathrm{ac}}$, ac resistivity $\rho$, and heat capacity $C$ measurements. The dc magnetization measurements were performed as a function of temperature $T$ at fixed field or as a function of applied magnetic field $\mu_0H$ at a fixed temperature in a Quantum Design Magnetic Property Measurement System (MPMS) magnetometer in temperatures ranging from 1.8 to 300~K and under magnetic fields up to 5~T. The ac susceptibility measurements were also performed in a Quantum Design MPMS with an ac applied field of 0.3~mT and a frequency of 30~Hz in dc magnetic fields up to 5~T. For field-dependent magnetization studies an Oxford Instruments vibrating sample magnetometer (VSM) was used with magnetic fields up to 10~T. Heat capacity was measured using a two-tau relaxation method in a Quantum Design Physical Property Measurement System (PPMS) at temperatures ranging from 1.9 to 300~K in magnetic fields up to 8~T. Lower-temperature measurements down to 0.5~K were carried out with a $^3$He insert. The samples were attached to the measuring stage using Apiezon\textsuperscript{\textregistered}~N grease to ensure good thermal contact. Electrical resistivity measurements were made using a conventional four-probe ac technique with a measuring frequency of 113~Hz and a current of 5.1~mA in a Quantum Design PPMS. The measurements were performed at temperatures ranging from 1.9 to 300~K in magnetic fields up to 9~T. The shape of the sample used for the majority of the measurements was a rectangular prism to allow the demagnetization factor to be evaluated~\cite{Aharoni1998} and minimized along one direction.

\section{Results and Discussion}

\subsection{Electrical resistivity}
\begin{figure}[tb!]
\centering
\includegraphics[width=0.8\columnwidth]{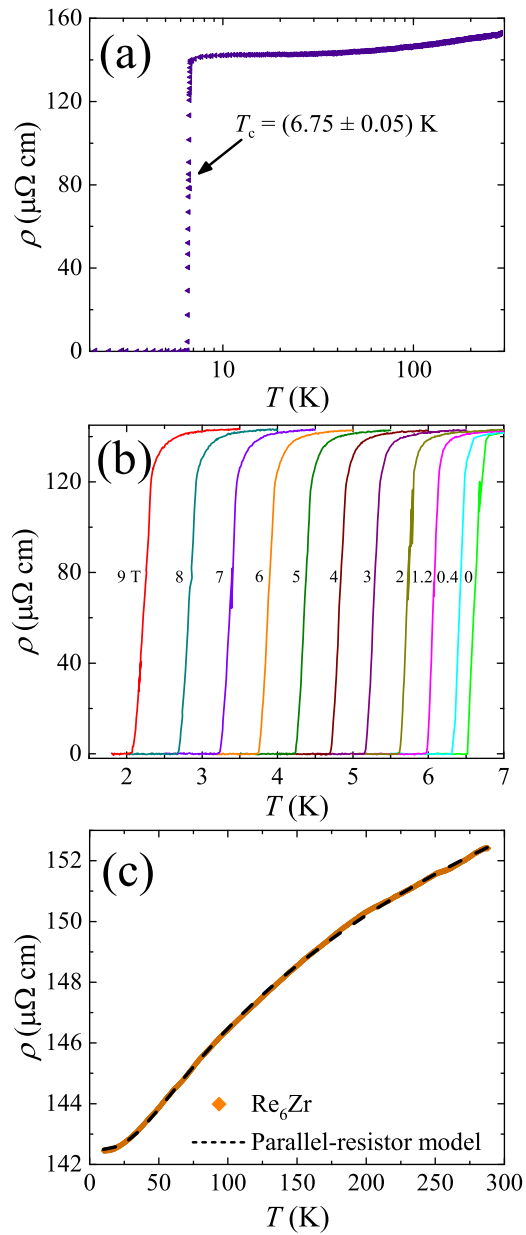}
\caption{\label{resistivity} (a) Resistivity versus temperature $\rho\left(T\right)$ of Re$_{6}$Zr in the range $1.8\leq  T\leq 250$~K measured in zero applied magnetic field. The midpoint of the resistivity drop was taken as the transition temperature. (b) $\rho\left(T\right)$ below 7.5 K shows the suppression of the transition temperature under various applied fields $\mu_0H$ from 0 to 9~T. (c) $\rho\left(T\right)$ data in the normal state fitted with the parallel-resistor model over the temperature range 10 to 290~K.}
\end{figure}

Figure~\ref{resistivity}(a) shows the resistivity as a function of temperature $\rho\left(T\right)$ of a polycrystalline Re$_{6}$Zr sample from 2 to 300~K in zero field. The small value of the residual resistivity ratio, RRR $\equiv \rho(300$~K$)/\rho(10$~K$) \approx 1.09 $, and the high normal-state resistivity at 10~K indicate poor metallic behavior. This is comparable to other Re compounds such as Re$_6$Hf with a RRR quoted from 1.08 to 1.4~\cite{Re6Hfa, Re6Hfb}, Re$_{24}$Ti$_5$ with RRR $\sim 1.3$~\cite{Re24Ti5}, and Nb$_{0.18}$Re$_{0.82}$ with  RRR $\sim 1.3$~\cite{NbRe}. A sharp, zero-field superconducting transition ($\Delta T_{\mathrm{c}}=0.20$~K) can be seen clearly in Fig.~\ref{resistivity}(b) at ${T_{\mathrm{c}} = \left(6.76\pm 0.05\right)}$~K. $T_{\mathrm{c}}$ is gradually suppressed with increasing applied magnetic field [see Fig.~\ref{resistivity}(b)] and the transition is broadened so that $\Delta T_{\mathrm{c}}=0.28$~K at 9~T.

At temperatures greater than $\sim 50$~K the $\rho\left(T\right)$ of Re$_{6}$Zr is seen to flatten. This characteristic is similar to that seen in many superconductors containing d-block elements including BiPd\cite{Joshi}. It has been proposed that in certain compounds at high temperatures the resistivity saturates at a value that corresponds to the mean free path on the order of the inter-atomic spacing\cite{Fisk}. This idea was further developed by Wiesmann $et~al.$\cite{Wiesmann} who found empirically that $\rho\left(T\right)$ could be described by the parallel-resistor model:
\begin{subequations}
\begin{equation}
\rho(T) = \left[\frac{1}{\rho_{\mathrm{sat}}} + \frac{1}{\rho_{\mathrm{ideal}}\left(T\right)}\right]^{-1}\label{PRMa},
\end{equation}
where $\rho_{\mathrm{sat}}$ is the saturated resistivity at high temperatures and is independent of $T$, and $\rho_{\mathrm{ideal}}(T)$ is the ``ideal'' contribution which according to Matthiessen's rule is: 
\begin{equation}
\rho_{\mathrm{ideal}}\left(T\right) = \rho_{\mathrm{ideal,}0} + \rho_{\mathrm{ideal,}L}\left(T\right)\label{PRMb}.
\end{equation}
Here $\rho_{\mathrm{ideal,}0}$ is the ideal temperature-independent residual resistivity and $\rho_{\mathrm{ideal,}L}\left(T\right)$ is the temperature-dependent contribution which can be expressed by the generalized Bloch-Gr\"uneisen model\cite{Grimvall}
\begin{multline}
\rho_{\mathrm{ideal,}L}\left(T\right) =C\left(\dfrac{T}{\Theta_{\mathrm{R}}}\right)^{n}\\
\times\int_{0}^{\Theta_{\mathrm{R}}/T} \dfrac{x^n}{\left(e^{x}-1\right)\left(1-e^{-x}\right)} dx\label{PRMc},
\end{multline}
\end{subequations}
where $\Theta_{\mathrm{R}}$ is the Debye temperature obtained from resistivity measurements, $C$ is a material-dependent pre-factor and $n=3-5$ depending on the nature of the carrier scattering. Fig.~\ref{resistivity}(c) shows the normal-state resistivity data from 10 to 290~K fit using Eq.~\ref{PRMa}. It was found that a value of $n=3$, which takes into account umklapp scattering between bands, achieved the best fit giving $\rho_{\mathrm{sat}}=\left(167\pm2\right)~\mathrm{\mu\Omega~cm}$, $C=\left(315\pm6\right)~\mathrm{\mu\Omega~cm}$  and $\Theta_{\mathrm{R}}=\left(237\pm2\right)$~K. The measured residual resistivity, $\rho_{0}~=~\left(142\pm2\right)~\mathrm{\mu\Omega~cm}$, which is related to  $\rho_{\mathrm{ideal,}0}$ and $\rho_{\mathrm{sat}}$ by 
\begin{equation}
\rho_{0} = \frac{\rho_{\mathrm{ideal,}0}\rho_{\mathrm{sat}}}{\rho_{\mathrm{ideal,}0} + \rho_{\mathrm{sat}}},
\end{equation}
is consistent with the values of the fit. This electrical resistivity data is in close agreement with that previously reported in Ref.~\onlinecite{Khan}.


\subsection{Heat capacity}

\begin{figure}[tb!]
\centering
\includegraphics[width=0.7\columnwidth]{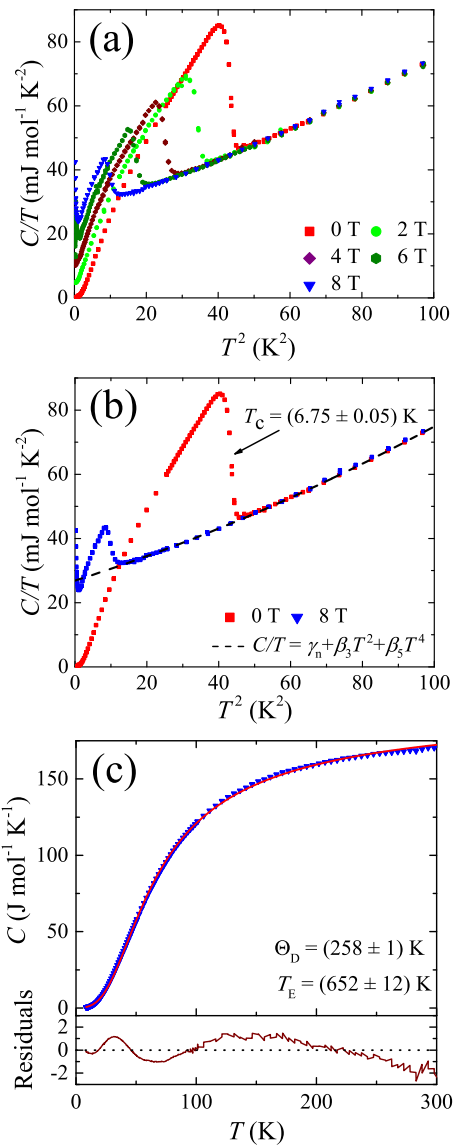}
\caption{\label{HC} (a) $C/T$ versus $T^{2}$ in different applied fields ($\mu_0H$ in teslas), showing the suppression in $T_{\mathrm{c}}$ for increasing field. (b) $C/T$ versus $T^{2}$ with $\mu_{0}H = 0$ and 8~T. The line is a fit using $C/T = \gamma_{\mathrm{n}} + \beta_{3}T^{2} + \beta_{5}T^{4}$ for all the $C\left(T\right)$ data collected above $T_{\mathrm{c}}\left(H\right)$ in the different applied fields. The normal-state electronic contribution to the specific heat ${\gamma_{\mathrm{n}}=\left(26.9\pm 0.1\right)\mathrm{mJ~mol}^{-1}~\mathrm{K}^{-2}}$, and the Debye temperature ${\Theta_{\mathrm{D}}=\left(338\pm 9\right)~\mathrm{K}}$. (c) $C$ versus $T$ from 10 to 300~K. The line shows the fit using Eq.~(\ref{DebyeEinstein}), the Debye-Einstein function. The residual plot underneath indicates the quality of the fit using the Debye-Einstein function to the data.}
\end{figure}

The temperature dependence of the heat capacity divided by temperature, $C/T$, versus $T^2$ from 2 to 10~K is shown in Fig.~\ref{HC}(a), where a sharp jump at $\left(6.75\pm 0.05\right)$~K indicates a bulk superconducting transition. The sharpness of this peak gives an indication of the high quality of the sample. We analyzed the normal-state data $C/T$ versus $T^2$ between 4.4 and 10~K at $\mu_{0}H = 0$~T using
\begin{equation} \label{Sommerfeld}
C/T = \gamma_{\mathrm{n}}+\beta_{3}T^{2}+\beta_{5}T^{4},
\end{equation}
where $\gamma_{\mathrm{n}}$ is the normal state Sommerfeld electronic-heat-capacity contribution, $\beta_{3}$ is the Debye law lattice-heat-capacity contribution, and $\beta_{5}$ is from highe- order lattice contributions. A fit using Eq.~(\ref{Sommerfeld}) can be seen in Fig.~\ref{HC}(b) which gives ${\gamma_{\mathrm{n}}=\left(26.9\pm 0.1\right)~\mathrm{mJ~mol}^{-1}~\mathrm{K}^{-2}}$, ${\beta_{3}=\left(0.35\pm 0.02\right)~\mathrm{mJ~mol}^{-1}~\mathrm{K}^{-4}}$ and ${\beta_{5}=\left(1.2\pm 0.3\right)~\mu\mathrm{J~mol}^{-1}~\mathrm{K}^{-6}}$. The Debye temperature, $\Theta_{\mathrm{D}}$, can then be calculated using
\begin{equation} \label{debye}
\Theta_{\mathrm{D}} = \left(\dfrac{12\pi^{4}RN}{5\beta}\right)^{1/3},
\end{equation}
where $R$ is the molar gas constant and $N$ is the number of atoms per unit cell. Eq.~\ref{debye} gives ${\Theta_{\mathrm{D}}=\left(338\pm 9\right)~\mathrm{K}}$ which is slightly higher than the previously reported value~\cite{Re6Zr}.

Figure~\ref{HC}(c) shows the temperature dependence of the heat capacity up to 300 K. There is no sign of any structural phase transition, and the value of $C$ at 300~K is ${169.5~\mathrm{J~mol}^{-1}~\mathrm{K}^{-1}}$, which is close to classical Dulong-Petit value for Re$_6$Zr of 174.6~J~mol$^{-1}~$K$^{-1}$ and  is consistent with $\Theta_{\mathrm{D}}>300$~K. We fit the normal-state data using a Debye-Einstein function. It was found that by including the additional Einstein term to the Debye model for lattice heat capacity the fit could be significantly improved. Figure~\ref{HC}(c) shows heat-capacity data from 10 to 300 K, which was fit with~\cite{Gopal}
\begin{subequations}
\begin{multline} \label{DebyeEinstein}
C(T) = \gamma_{\mathrm{n}}T + n \delta C_{\mathrm{Debye}}\left(\frac{T}{\Theta_{\mathrm{D}}}\right)\\
+ n (1-\delta) C_{\mathrm{Einstein}}\left(\frac{T}{T_{\mathrm{E}}}\right),
\end{multline}
where $\delta$ is the fractional contribution of $C_{\mathrm{Debye}}$, $n$ is the number of atoms in a formula unit (f.u.), $C_{\mathrm{Debye}}$ is given by
\begin{equation}
C_{\mathrm{Debye}}\left(\dfrac{T}{\Theta_{\mathrm{D}}}\right) = 9R\left(\dfrac{T}{\Theta_{\mathrm{D}}}\right)^{3}
\int_{0}^{\Theta_{\mathrm{D}}/T}\dfrac{x^{4}e^{x}}{\left(e^{x}-1\right)^{2}}dx,
\end{equation}
and $C_{\mathrm{Einstein}}$ is given by
\begin{equation}
C_{\mathrm{Einstein}}\left(\frac{T}{T_{\mathrm{E}}}\right) = 3R \dfrac{z^{2}e^{z}}{\left(e^{z}-1\right)^{2}},
\end{equation}
\end{subequations} 
where $z = T_{\mathrm{E}}/T$ and $T_{\mathrm{E}}$ is the Einstein temperature. The fit was performed using a fixed value ${\gamma_{\mathrm{n}}=26.9~\mathrm{mJ}~\mathrm{mol}^{-1}\mathrm{K}^{-2}}$ to help reduce the number of free parameters. We obtained $\delta = 0.912 \pm 0.002 $, $\Theta_{\mathrm{D}} = \left(258\pm 1\right)$~K, and $T_{\mathrm{E}} = \left(652 \pm 12\right)$~K. The difference between $\Theta_{\mathrm{D}}$ and $\Theta_{\mathrm{R}}$ is also expected due to the limitations of the parallel-resistor model.


In Fig.~\ref{HC}, at very low temperatures, an upturn in $C/T$ appears in magnetic fields above 6~T. This anomalous contribution to the specific heat is proportional to $T^{-2}$, which suggests that it is due to the high-temperature tail of a nuclear Schottky anomaly. The specific heat of the measured Re$_{6}$Zr can be expressed as
\begin{equation}
C (T,B) = C_{\mathrm{el}}(T,B) + C_{\mathrm{ph}}(T) + C_{\mathrm{hf}}(T,B),
\end{equation}
where $C_{\mathrm{el}}$ is the electronic contribution, $C_{\mathrm{hf}}$ is the Schottky contribution, and $C_{\mathrm{ph}}$ is the phonon contribution. The high-temperature approximation of the nuclear hyperfine contribution to the specific heat was modeled by $C_{\mathrm{hf}} = A_{0} T^{-2}$, where $A_{0}$ is a field-dependent parameter. $A_{0}$ is estimated to be ${\sim 1.4}$~mJ K mol$^{-1}$ at 8~T, which is consistent with the value previously obtained for pure rhenium~\cite{ReHF, ReHF2}. The results of this analysis raise a note of caution. 

A hyperfine contribution to the specific heat has also been seen in other Re-based $\alpha$-Mn compounds, Nb$_{0.18}$Re$_{0.82}$~\cite{NbRea} and Re$_{6}$Hf~\cite{Re6Hfb}, as well as in pure Re~\cite{ReHF, ReHF2}, indicating that a Schottky anomaly may always be present in Re-based superconductors at low temperatures. Mazidian~\textit{et al.} demonstrated that in order to establish the presence of point or line nodes in the superconducting gap, the heat capacity needs to be fit below $T_{\mathrm{c}}/10$~\cite{Mazidian}. Modifications by a magnetic field below $T_{\mathrm{c}}$ to both $C_{\mathrm{el}}(T,B)$ and $C_{\mathrm{hf}}(T,B)$ mean that a precise evaluation of the temperature dependence of the electronic specific heat and hence the gap structure in all Re-based NCS superconductors, including those with an $\alpha$-Mn structure, may be challenging, as this will require an accurate evaluation of the hyperfine contribution to the specific heat.

\subsection{Magnetization and lower critical field}

\begin{figure}[tb!]
\centering
\includegraphics[width=0.7\columnwidth]{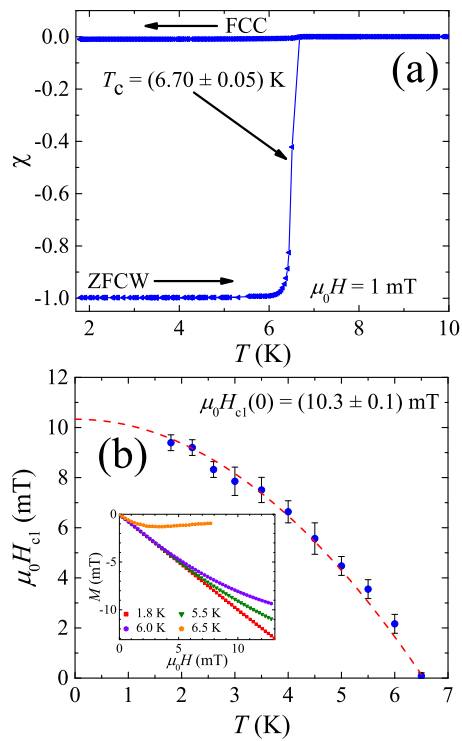}
\caption{\label{HC1} (a) Temperature dependence of the dc magnetic susceptibility $\chi_{\mathrm{dc}}\left(T\right)$ collected in zero-field-cooled warming (ZFCW) and field-cooled cooling (FCC) modes in an applied field of $\mu_{0}H=1$~mT. (b) Lower critical field $H_{\mathrm{c1}}$  versus temperature for Re$_{6}$Zr. The $H_{\mathrm{c1}}$ values were taken as the fields at which initial magnetization versus field data shown in Fig.~\ref{HC1}(b) first deviate from linearity (as shown in the inset). The solid line shows the fit using Eq.~(\ref{GLHc1}) giving $\mu_{0}H_{\mathrm{c1}}(0)=\left(10.3\pm 0.1\right)$~mT.}
\end{figure}

Figure~\ref{HC1}(a) shows the dc susceptibility data $\chi_{\mathrm{dc}}\left(T\right)$ taken in zero-field-cooled warming (ZFCW) and field-cooled cooling (FCC) modes in an applied field of 1~mT. These data confirm that Re$_{6}$Zr is a superconductor with $T_{\mathrm{c}}=\left(6.70\pm 0.05\right)$~K. The sample exhibits a full Meissner fraction for the ZFCW. There is almost no flux expulsion on re-entering the superconducting state during FCC. The strong pinning is consistent with a disordered system. Magnetization versus field sweeps in low fields (0 to 16~mT) at several temperatures are shown in Fig.~\ref{HC1}(b). The lower critical field, $H_{\mathrm{c1}}\left(T\right)$, is determined from the first deviation from linearity of the initial slope as the field is increased. In Fig.~\ref{HC1}(c) the resulting $H_{\mathrm{c1}}\left(T\right)$ values are plotted against temperature. Ginzburg-Landau (GL) theory gives
\begin{equation} \label{GLHc1}
H_{\mathrm{c1}}(T) = H_{\mathrm{c1}}(0)\left[1-\left(\dfrac{T}{T_{\mathrm{c}}}^2\right)\right].
\end{equation}
Fitting the data using Eq.~(\ref{GLHc1}), $H_{\mathrm{c1}}\left(0\right)$ was estimated to be $\left(10.3\pm 0.1\right)$~mT.

\begin{figure}[tb!]
\centering
\includegraphics[width=0.7\columnwidth]{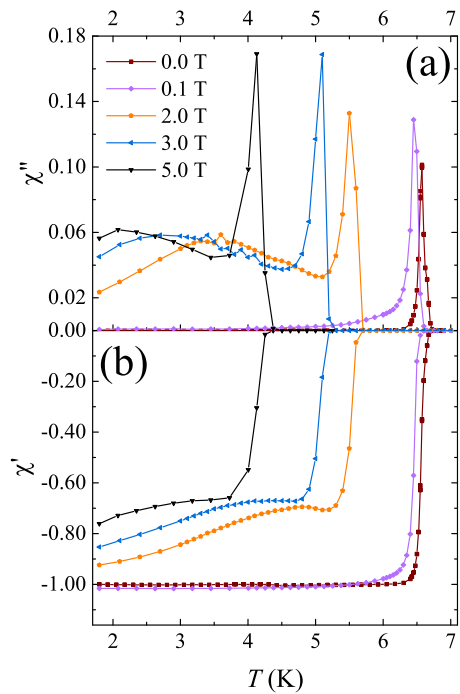}
\caption{\label{ACtrans} (a) Imaginary part of ac susceptibility versus temperature $\chi''\left(T\right)$ in various dc applied fields. (b) Real part of ac susceptibility versus temperature $\chi'\left(T\right)$ at various dc applied fields. In zero dc field, a sharp superconducting transition can be seen at $\left(6.70\pm 0.05\right)$~K. In fields above $H_{\mathrm{c1}}\left(0\right)$ an anomalous dip in the magnetization is seen close to the transition temperature.}
\end{figure} 

The ac susceptibility versus temperature measurements $\chi_{\mathrm{ac}}\left(T\right)$ shown  in Fig.~\ref{ACtrans} confirm the superconducting transition of $T_{\mathrm{c}}=\left(6.70\pm 0.05\right)$~K. In dc bias fields less than $H_{\mathrm{c1}}\left(0\right)$ the sample exhibits a full Meissner fraction. The out-of-phase component of the ac susceptibility $\chi''(T)$ contains a sharp maximum close to $T_{\mathrm{c}}$ and falls to zero for lower temperatures. This is consistent with the strong flux pinning seen in the low-field FCC $M\left(T\right)$ data.  For applied fields much greater than $H_{\mathrm{c1}}\left(0\right)$, $T_{\mathrm{c}}$ is suppressed, and a full Meissner fraction is not seen due to partial flux penetration. An anomalous dip can be seen close to $T_{\mathrm{c}}$, suggesting flux is being reexpelled from the sample due to unusual flux dynamics. At lower temperature, $\chi''(T)$ exhibits a broad maximum, indicating losses due to flux motion in dc applied fields $\mu_0H\geq 2$~T. 

\begin{figure}[tb!]
\centering
\includegraphics[width=0.7\columnwidth]{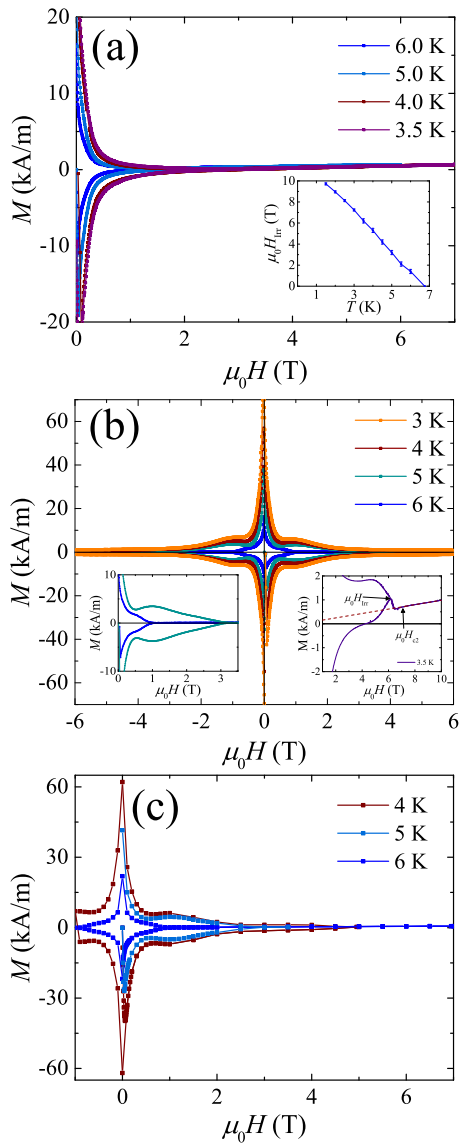}
\caption{\label{MvH} (a) Magnetization vs magnetic field at several temperatures for Re$_{6}$Zr. The data were collected in a VSM on a plate-shaped sample with the demagnetization factor of the sample minimized. The inset shows how $H_{\mathrm{Irr}}$ varies with temperature.  (b) Magnetization vs. magnetic field at several temperatures collected in a vibrating sample magnetometer with the demagnetization factor of the Re$_{6}$Zr sample maximized. A secondary maximum (fishtail) can clearly be seen in the magnetization at around 1.25~T. The left inset shows the 5 and 6~K curves between 0 and 3.5~T. $H_{\mathrm{Irr}}$ and $H_{\mathrm{c}2}$ are indicated in the right inset showing the 3.5~K curve between 2 and 10~T. (c) Magnetization vs magnetic field at several temperatures collected in the SQUID magnetometer. The fishtail can also be clearly seen in a magnetic field of $\sim 1.25$~T.}
\end{figure}

Further evidence of unusual flux pinning in Re$_{6}$Zr can be seen in the $M\left(H\right)$ loops taken in the both the superconducting quantum interference device (SQUID) magnetometer and the VSM (see Fig.~\ref{MvH}), suggesting that the observed features cannot simply be attributed to the significant movement of the sample in a magnetic field or the magnetic field sweep rate. As is evident from Fig.~\ref{MvH}(a), above $H_{\mathrm{c1}}$, Re$_{6}$Zr exhibits the conventional behavior for a type-II superconductor, with a hysteresis in the magnetization $\Delta M$ decreasing with increasing temperature and magnetic field. For applied fields close to $H_{\mathrm{c2}}\left(T\right)$ this hysteresis $\Delta M$ disappears, and the magnetization becomes reversible as vortices appear to become unpinned. The inset in Fig.~\ref{MvH}(a) shows how this irreversibility field $H_{\mathrm{Irr}}$ varies with temperature. These data were collected using a plate-shaped sample with the field applied in the plane of the plate, i.e., with the demagnetization factor of the sample minimized. By changing the sample orientation with respect to the applied field a change in vortex pinning is observed, as can be seen in Figs.~\ref{MvH}(b) and ~\ref{MvH}(c), where the demagnetization factor was maximized. In Figs.~\ref{MvH}(b) and ~\ref{MvH}(c) a clear secondary maximum (fishtail) is observed. As the sample is cooled, there is a slight shift to higher magnetic field in the onset and the peak of the fishtail. This behavior is not normally observed in low-$T_{\mathrm{c}}$ superconductors but is quite common in the high-$T_{\mathrm{c}}$ oxides and in some two-dimensional superconducting materials, indicating unconventional vortex states. The symmetry of the hysteresis in the field-increasing and field-decreasing legs of the $M\left(H\right)$ curves suggests that bulk pinning rather than surface barriers may be the dominant mechanism leading to the fishtail. Assuming the superconducting critical current is proportional to $\Delta M$, the maximum pinning force in the field range  1 to 3~T, as reflected in the fishtail, appears to be almost temperature independent between 3 and 5~K. It is suggested that the unusual vortex states arise from the normal pinning centers such as grain boundaries within the sample. A detailed study on the vortex states in high-quality single crystals of Re$_{6}$Zr is needed to explore the vortex physics further.

\subsection{Superconducting gap}
\label{sec:Superconducting Gap}

\begin{figure}[tb!]
\centering
\includegraphics[width=0.7\columnwidth]{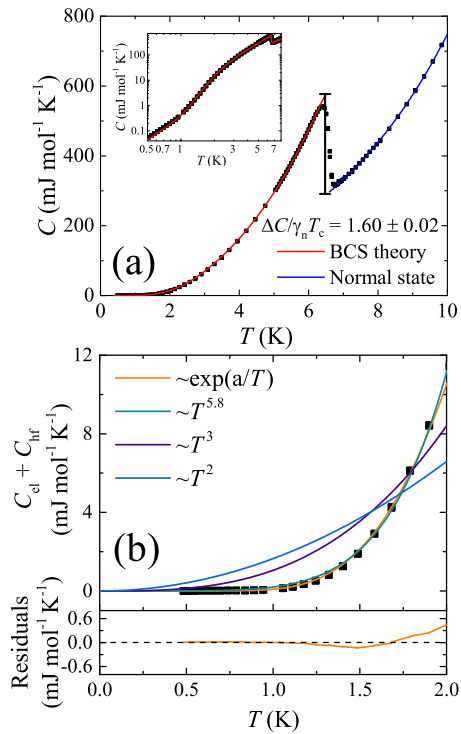}
\caption{\label{SCgap} (a) Heat capacity versus temperature in zero field for Re$_{6}$Zr across the superconducting transition. The red line shows a fit across the superconducting transition for a fully gapped superconductor as described in Sec.~\ref{sec:Superconducting Gap}. The inset shows the heat capacity across the superconducting transition on a log-log scale. From this it can be see that the data are very well fit by the isotropic $s$-wave BCS model.  (b) Electronic heat capacity $C_{\mathrm{el}}$ versus temperature below 2.5~K showing various power laws (anisotropic gap) and an exponential (isotropic gap) fit to the low-temperature data. The $\chi^{2}$ and residuals shown are for the exponential fit.}
\end{figure}

The jump in specific heat in zero field indicates the onset of bulk superconductivity. The transition temperature is defined as the midpoint of the transition, giving $T_{\mathrm{c}}=\left(6.75\pm 0.05\right)$~K. The data in Fig.~\ref{SCgap}(a) were fit using the BCS model of the specific heat given in Ref.~\onlinecite{Gapfitting}. The entropy $S$ was calculated from
\begin{equation}
\dfrac{S}{\gamma_{\mathrm{n}} T_{\mathrm{c}}}=-\dfrac{6}{\pi^2}\dfrac{\Delta_{0}}{k_{\mathrm{B}}T_{\mathrm{c}}}\int_{0}^{\infty}\left[f\mathrm{ln}f+\left(1-f\right)\mathrm{ln}\left(1-f\right)\right]dy,
\end{equation} 
where $f$ is the Fermi-Dirac function given by ${f=\left[1+\exp\left(E/k_{\mathrm{B}}T\right)\right]^{-1}}$ and ${E=\Delta_{0}\sqrt{y^{2}+\delta(T)^{2}}}$, where $y$ is the energy of the normal-state electrons and $\delta(T)$ is the temperature dependence of the superconducting gap calculated from the BCS theory. The specific heat of the superconducting state is then calculated by
\begin{equation}
\dfrac{C_{\mathrm{sc}}}{\gamma_{\mathrm{n}} T_{\mathrm{c}}}=T\dfrac{d(S/\gamma_{\mathrm{n}} T_{\mathrm{c}})}{dT}.
\end{equation}
The superconducting gap $\Delta_{0}/k_{\mathrm{B}}T_{\mathrm{c}}$ was estimated to be $1.86\pm 0.05$, which is in agreement with Ref.~\onlinecite{Re6Zr}. For conventional BCS superconductors a value of 1.76 is expected, and the larger value for Re$_{6}$Zr indicates that the electron-phonon coupling is slightly enhanced. ${\Delta C/\gamma_{\mathrm{n}} T_{\mathrm{c}} = 1.60 \pm 0.02}$ is also larger than the 1.43 expected for conventional BCS superconductors and agrees with the values reported in Refs.~\onlinecite{Re6Zr} and~\onlinecite{Khan}. A fit was also attempted using a two-gap model, but it was found that $\Delta_{0}/k_{\mathrm{B}}T_{\mathrm{c}}$ for the two gaps iterated to the same value, indicating that the material has a single gap.

To determine whether the superconducting gap is isotropic (exponential) or anisotropic (power law) it is necessary to determine the temperature dependence of the electronic component of the heat capacity down to low temperature, as shown in Fig.~\ref{SCgap}(b). Due to the difficulties in approximating the zero-field hyperfine contribution in the specific heat this contribution has also been included in Fig.~\ref{SCgap}(b). Figure~\ref{SCgap}(b) shows fits to several power laws of the form $~b \times T^{N}$, where $b$ is a constant. Setting $N~=~2$ or 3 the fits are poor, while $N~=~5.8$ gives a good fit to the data, although this provides no physical insight. The $\left(C_{\mathrm{el}} + C_{\mathrm{hf}}\right)$ data are best described by an exponential temperature dependence, suggesting an isotropic fully gapped $s$-wave BCS superconductor. To obtain the true nature of the superconducting gap heat-capacity data well below $T_{\mathrm{c}}/10$ need to be analyzed~\cite{Mazidian}. From Fig.~\ref{SCgap}(a) it can be seen that the specific heat is rather low. A more complete understanding of the hyperfine term is required to make any further progress with this analysis. Nuclear quadrupole measurements have also been performed on Re$_{6}$Zr and provide further evidence of a conventional BCS gap symmetry \cite{Matano}.

\subsection{Upper critical field}

\begin{figure}[tb!]
\centering
\includegraphics[width=0.7\columnwidth]{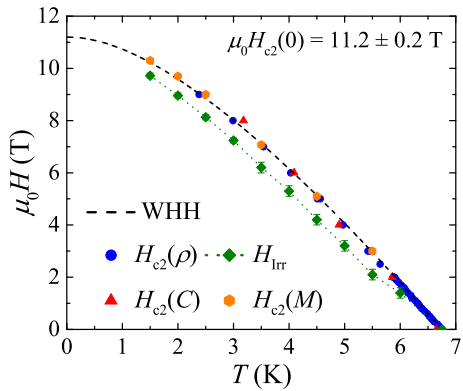}
\caption{\label{WHHfit} Upper critical field versus temperature of Re$_6$Zr determined from the electrical resistivity, heat capacity, and magnetization data. The black curve shows the prediction for $H_{\mathrm{c}2}\left(T\right)$ from the WHH model. For comparison $H_{\mathrm{Irr}}\left(T\right)$ from Fig.~\ref{MvH}(b) is included. $H_{\mathrm{Irr}}\left(T\right)$ can be seen to diverge away from $H_{\mathrm{c}2}\left(T\right)$ close to $T_{\mathrm{c}}$ and then stay a constant distant from $H_{\mathrm{c}2}\left(T\right)$ down to 1.5~K.}
\end{figure}

In order to measure the upper critical field as a function of temperature $H_{\mathrm{c}2}\left(T\right)$, the shift in $T_{\mathrm{c}}$ in magnetic fields of up to 9~T was determined from heat-capacity and resistivity data. 

Figure~\ref{WHHfit} shows how $H_{\mathrm{c}2}$ varies with $T$. At temperatures just below $T_{\mathrm{c}}$ it is clear that $H_{\mathrm{c}2}$ increases linearly with decreasing $T$, and this indicates that the temperature dependence of $H_{\mathrm{c}2}$ given by the Ginzburg-Landau formula is not appropriate. Instead, the Werthamer-Helfand-Hohenberg (WHH) model was used. This allows $H_{\mathrm{c}2}\left(0\right)$ to be calculated in terms of the spin-orbit scattering and Pauli spin paramagnetism~\cite{WHH}, as it is expected that spin-orbit coupling may be strong due to the presence of the rhenium. $H_{\mathrm{c}2}\left(T\right)$ can be found by solving 
\begin{multline}
\ln\left(\dfrac{1}{t}\right) = \left(\dfrac{1}{2} + \dfrac{i\lambda_{\mathrm{so}}}{4\gamma}\right)\psi\left(\dfrac{1}{2}+\dfrac{\bar{h}+\frac{1}{2}\lambda_{\mathrm{so}}+i\gamma}{2t}\right)\\
+\left(\dfrac{1}{2}-\dfrac{i\lambda_{\mathrm{so}}}{4\gamma}\right)\psi\left(\dfrac{1}{2}+\dfrac{\bar{h}+\frac{1}{2}\lambda_{\mathrm{so}}+i\gamma}{2t}\right)-\psi\left(\dfrac{1}{2}\right),
\end{multline}
where $t = T/T_{\mathrm{c}}$, $\lambda_{\mathrm{so}}$ is the spin-orbit scattering parameter, $\alpha_{\mathrm{M}}$ is the Maki parameter, $\psi$ is the digamma function, $\bar{h}$ is the dimensionless form of the upper critical field given by
\begin{equation}
\bar{h} = \dfrac{4 H_{\mathrm{c}2}}{\pi^2}\left(\dfrac{dH_{\mathrm{c}2}}{dT}\right)^{-1}_{t=1},
\end{equation}
and $\gamma=\sqrt{(\alpha\bar{h})^2-(\frac{1}{2} \lambda_{\mathrm{so}})^2}$. It is estimated that $\mu_{0}H_{\mathrm{c2}}\left(0\right) = \left(11.2 \pm 0.2\right)$~T, close to the value reported by Ref.~\citen{Re6Zr} but below the Pauli paramagnetic limiting field $\mu_{0}H_{\mathrm{Pauli}}$ of ${\left(12.35\pm 0.09\right)~\mathrm{T}}$.

The WHH expression has three variables: the Maki parameter $\alpha_{\mathrm{M}}$, the spin-orbit scattering parameter $\lambda_{\mathrm{so}}$, and the gradient at $T_{\mathrm{c}}$. In their original work~\cite{WHH}, WHH state that $\alpha_{\mathrm{M}}$ is not an adjustable parameter and needs to be obtained from experimental data; thus, $\alpha_{\mathrm{M}}$ was not varied during the fitting. 

The Maki parameter can be estimated experimentally by using the WHH expression 
\begin{equation} \label{maki}
\alpha_{\mathrm{M}} = \sqrt{2}\dfrac{H^{\mathrm{orb}}_{\mathrm{c2}}\left(0\right)}{H^{\mathrm{Pauli}}_{\mathrm{c2}}\left(0\right)},
\end{equation}
where $H^{\mathrm{orb}}_{\mathrm{c2}}$ is the orbital limiting field given by
\begin{equation}
H^{\mathrm{orb}}_{\mathrm{c2}}\left(0\right) = -\alpha T_{\mathrm{c}}\dfrac{-dH_{\mathrm{c2}}(T)}{dT}\bigg\vert_{T=T_{\mathrm{c}}},
\end{equation}
where $\alpha$ is the purity factor, which for superconductors in the dirty limit is 0.693. The initial slope $-dH_{\mathrm{c2}}(T)/dT\vert_{T=T_{\mathrm{c}}}$ was found to be 2.44~T/K, giving ${\mu_{0}H^{\mathrm{orb}}_{\mathrm{c2}}\left(0\right)=\left(11.41\pm 0.05\right)~\mathrm{T}}$. From Eq.~(\ref{maki}) we obtain $\alpha_{\mathrm{M}}=1.31$, and the relative size of the Maki parameter indicates that the Pauli limiting field is non-negligible. 
Fixing $\alpha_{\mathrm{M}}= 1.31$ produced ${\lambda_{\mathrm{so}}=18\pm 5}$. It was found that this model is highly dependent on the starting values as an equally good fit, as judged by the reduced $\chi^2$, could be obtained by allowing the Maki parameter to vary. $\alpha_{\mathrm{M}}$ was found to drift towards zero in nearly all cases along with $\lambda_{\mathrm{so}}$, which would also tend to zero when allowed to vary. Unsurprisingly, the initial gradient $-dH_{\mathrm{c2}}(T)/dT\vert_{T=T_{\mathrm{c}}}$ was found to remain constant within error. 

In the first case with $\alpha_{\mathrm{M}}$ fixed, the value for the spin-orbit term seems unusually large. There are several reasons why the WHH model may misrepresent what is happening in the system: (1) \textit{Two-gap superconductor.} While the analysis of the superconducting gap was assumed to be a single gap it is possible that Re$_6$Zr is a two-gap superconductor where the gaps are of a similar magnitude, and this would give rise to some enhancement of $H_{\mathrm{c}2}$~\cite{Gurevich}. (2) \textit{Granularity}. The polycrystalline sample of Re$_{6}$Zr will contain grain boundaries. The upper critical field will be increased above the bulk value once the grain size becomes smaller than the coherence length~\cite{deGennes} (the grain size is unknown, so contributions from this source are unclear). (3) \textit{Spin-orbit coupling}. Strong spin-orbit coupling effects can yield large enhancements of $H_{\mathrm{c}2}$ such that the temperature dependence of $H_{\mathrm{c}2}$ can become linear, although in the dirty limit this enhancement should be weaker~\cite{Carbotte}. In order to obtain a more accurate value for $\mu_{0}H_{\mathrm{c}2}\left(0\right)$ high-field, low-temperature measurements of $H_{\mathrm{c}2}$ are needed in order to determine the form of the $\mu_{0}H_{\mathrm{c}2}\left(T\right)$ curve much closer to ${T=0~\mathrm{K}}$.

\subsection{Properties of the superconducting state}

The results of resistivity, heat-capacity, and magnetization measurements can now be combined in order to estimate some of the important superconducting properties of Re$_6$Zr. The Ginzburg-Landau coherence length $\xi_{\mathrm{GL}}(0)$ can be estimated using $\mu_{0}H_{\mathrm{c}2}\left(0\right)$ from~\cite{Poole}
\begin{equation}
H_{\mathrm{c}2}\left(0\right) = \dfrac{\Phi_{0}}{2\pi\xi^{2}_{\mathrm{GL}}(0)},
\end{equation}
where $\Phi_{0} = 2.07 \times 10^{-15}$~Wb is the magnetic flux quantum. We calculate $\xi_{\mathrm{GL}}(0) = \left(5.37~\pm~0.09\right)$~nm. $\mu_{0}H_{\mathrm{c}1}\left(0\right)$ and $\xi_{\mathrm{GL}}(0)$ can be used to calculate the Ginzburg-Landau penetration depth $\lambda_{\mathrm{GL}}\left(0\right)$ from the relation
\begin{equation}\label{hc1lambda}
H_{\mathrm{c}1}\left(0\right) = \left(\dfrac{\Phi_{0}}{4\pi\lambda^{2}_{\mathrm{GL}}(0)}\right)\ln\left(\dfrac{\lambda_{\mathrm{GL}}(0)}{\xi_{\mathrm{GL}}(0)}\right).
\end{equation}
Using $\mu_{0}H_{\mathrm{c}1} = 10.3$ mT and $\xi_{\mathrm{GL}}(0) = 5.37$~nm, we calculated ${\lambda_{\mathrm{GL}(0)}=\left(247\pm 4\right)}$~nm. The Ginzburg-Landau parameter can now be calculated by the relation
\begin{equation}
\kappa_{\mathrm{GL}} = \dfrac{\lambda_{\mathrm{GL}}(0)}{\xi_{\mathrm{GL}}(0)}.
\end{equation}
This yields a value of $\kappa_{\mathrm{GL}}=46.2\pm 0.8$. For a superconductor to be classed as a type-II superconductor $\kappa_{\mathrm{GL}}\geq\frac{1}{\sqrt{2}}$. It is clear that Re$_{6}$Zr is a strong type-II superconductor.

The thermodynamic critical field $H_{\mathrm{c}}$ can be calculated using $\xi_{\mathrm{GL}}(0)$ and $\lambda_{\mathrm{GL}}(0)$ using the relation
\begin{equation}
H^{\mathrm{cal}}_{\mathrm{c}}\left(0\right) = \dfrac{\Phi_{0}}{2\sqrt{2}\pi\xi_{\mathrm{GL}}(0)\lambda_{\mathrm{GL}}(0)},
\end{equation}
from this $H^{\mathrm{cal}}_{\mathrm{c}}\left(0\right)$ is estimated to be $\left(175 \pm 3\right)$~mT. The thermodynamic critical field can be experimentally estimated from the difference between the free energies per unit volume of the superconducting and normal states $\Delta F$ by~\cite{Poole}
\begin{equation}\label{thermcrit}
\frac{H^{2}_{\mathrm{c}}(T)}{8 \pi} = \Delta F = \int_{T_{\mathrm{c}}}^{T} \int_{T_{\mathrm{c}}}^{T'} \frac{C_{\mathrm{s}}-C_{\mathrm{n}}}{T''} dT'' dT',
\end{equation}
where $C_{\mathrm{s}}$ and $C_{\mathrm{n}}$ are the heat capacities per unit volume. From Eq.~(\ref{thermcrit}) we obtain $H^{\mathrm{exp}}_{\mathrm{c}}\left(0\right) = \left(130 \pm 2\right)$~mT. 

In order to calculate the electronic mean free path and London penetration depth in Re$_{6}$Zr the Sommerfeld coefficient can be written as~\cite{kittel}
\begin{equation}\label{gamman}
\gamma_{n}=\left(\dfrac{\pi}{3}\right)^{2/3} \dfrac{k_{\mathrm{B}}^{2} m^{*} V_{\mathrm{f.u.}} n^{1/3}}{\hbar^{2} N_{\mathrm{A}}},
\end{equation}
where $k_{\mathrm{B}}$ is the Boltzmann constant, $N_{\mathrm{A}}$ is the Avogadro constant, $V_{\mathrm{f.u.}}$ is the volume of a formula unit, $m^{*}$ is the effective mass of quasiparticles, and $n$ is the quasiparticle number density per unit volume. The electronic mean free path $\ell_{\mathrm{e}}$ can be estimated from the residual resistivity $\rho_{\mathrm{0}}$ by the equation
\begin{equation}\label{meanfreepath}
\ell_{\mathrm{e}} = \dfrac{3 \pi^{2} \hbar^{3}}{e^{2} \rho_{\mathrm{0}} m^{*2} \nu^{2}_{\mathrm{F}}},
\end{equation}
where the Fermi velocity $\nu_{\mathrm{F}}$ is related to the effective mass and the carrier density by
\begin{equation}\label{n}
n = \dfrac{1}{3 \pi^{2}}\left( \dfrac{m^{*} \nu_{\mathrm{F}}}{\hbar} \right)^{3}.
\end{equation}
In the dirty limit the penetration depth is given by
\begin{equation}\label{lambda0}
\lambda_{\mathrm{GL}}(0) = \lambda_{\mathrm{L}}\left( 1 + \dfrac{\xi_{\mathrm{0}}}{\ell_{\mathrm{e}}} \right)^{1/2},
\end{equation}
where $\xi_{\mathrm{0}}$ is the BCS coherence length and $\lambda_{\mathrm{L}}$ is the London penetration depth, which is given by
\begin{equation}\label{lambdaL}
\lambda_{\mathrm{L}} = \left( \dfrac{m^{*}}{\mu_{0} n e^{2} } \right)^{1/2}.
\end{equation}
The Ginzburg-Landau coherence length is also affected in the dirty limit. The relationship between the BCS coherence length $\xi_{\mathrm{0}}$ and the Ginzburg-Landau coherence $\xi_{\mathrm{GL}}$ at $T~=~0$ is
\begin{equation}\label{xi}
\dfrac{\xi_{\mathrm{GL}}(0)}{\xi_{0}} = \dfrac{\pi}{2\sqrt{3}}\left(1 + \dfrac{\xi_{\mathrm{0}}}{\ell_{\mathrm{e}}} \right)^{-1/2}. 
\end{equation}
Equations~(\ref{gamman})~-~(\ref{xi}) form a system of four equations. To estimate the parameters $m^{*}$, $n$, $\ell_{\mathrm{e}}$, and $\xi_{0}$ this system of equations can be solved simultaneously using the values $\gamma_{\mathrm{n}}~=~26.9~\mathrm{mJ~mol^{-1}~K^{-2}}$, $\xi_{\mathrm{GL}}~=~5.37$~nm, and $\rho_{0}~=~142\mathrm{\mu\Omega~cm}$. For comparison, two values of $\lambda_{\mathrm{GL}}$ have been used; $247$~nm is taken from Eq.~(\ref{hc1lambda}), and $356$~nm is taken from the $\mu$SR study in Ref.~\onlinecite{Re6Zr}. The results are shown in Table~\ref{Elecprop}. From the mean free path $\ell_{\mathrm{e}}$ calculated in Eq.~(\ref{meanfreepath}) and $\xi_0$ calculated in Eq.~(\ref{xi}), it is clear that $\xi_0 > \ell_{\mathrm{e}}$, indicating that Re$_{6}$Zr is in the dirty limit. We find that these values are in close agreement with those previously reported for Re$_{6}$Zr~\cite{Khan}. 

\begin{table}[tb!]
\caption{Comparison of electronic properties of Re$_{6}$Zr for $\lambda_{\mathrm{GL}} \left(H_{\mathrm{c1}}\right)$ and $\lambda_{\mathrm{GL}} \left(\mathrm{\mu SR}\right)$.}\label{Elecprop}
\begin{center}
 \begin{tabular}{l l r r} 
 \hline\hline
  Property & Units & $H_{\mathrm{c1}} $ & $\mu$SR \\ 
 \hline
 & & & \\
 
 $\lambda_{\mathrm{GL}}(0)$ & nm & $247$ & $356$ \\ 

 $m^*/m_{\mathrm{e}}$ & & $10.1 \pm 0.1$ & $12.9 \pm 0.02$ \\
 
 $m^{*}_{\mathrm{band}}/m_{\mathrm{e}}$ & & $6.0 \pm 0.1$ & $7.7 \pm 0.1$ \\
 
 $n$ & $10^{27}$m$^{-3}$ & $15.2 \pm 0.2$ & $7.4 \pm 0.1$ \\

 $\xi_{\mathrm{0}}$ & nm & $3.28 \pm 0.5$ & $3.70 \pm 0.05$ \\
 
 $\ell_{\mathrm{e}}$ & nm & $1.45 \pm 0.02$ & $2.36 \pm 0.03$ \\
 
 $\xi_{\mathrm{0}} / \ell_{\mathrm{e}}$ & & $2.25 \pm 0.03$ & $1.56 \pm 0.02$ \\
 
 $\lambda_{\mathrm{L}}$ & nm & $136 \pm 2$ & $222 \pm 3$ \\
 
 $\nu_{\mathrm{F}}$ & m s$^{-1}$ & $88000 \pm 1000$ & $54000 \pm 800$ \\
 
 $T_{\mathrm{F}}$ & K & $2570 \pm 40$ & $1240 \pm 20$ \\
 
 $T_{\mathrm{c}} / T_{\mathrm{F}}$ & & $0.0026 \pm 0.0001 $ & $0.0054 \pm 0.0001$ \\

 & & & \\
 \hline\hline
\end{tabular}
\end{center}
\end{table}
  
The bare-band effective mass $m^{*}_{\mathrm{band}}$ can be related to $m^*$, which contains enhancements from the many-body electron-phonon interactions~\cite{Grimvall1976}
\begin{equation}\label{mband}
m^{*} = m^{*}_{\mathrm{band}}\left(1+\lambda_{\mathrm{el-ph}}\right),
\end{equation}
where $\lambda_{\mathrm{el-ph}}$ is the electron-phonon coupling constant. The electron-phonon coupling constant gives the strength of the interaction between electron and phonons in superconductors. This can be estimated from McMillan's theory~\cite{Mcmillan1968} from $\Theta_{\mathrm{D}}$ and $T_{\mathrm{c}}$,
\begin{equation}
\lambda_{\mathrm{el-ph}}=\dfrac{1.04 + \mu^{*}\ln\left(\Theta_{\mathrm{D}}/1.45T_{\mathrm{c}}\right)}{\left(1-0.62\mu^{*}\right)\ln\left(\Theta_{\mathrm{D}}/1.45T_{\mathrm{c}}\right)-1.04},
\end{equation}
where $\mu^{*}$ is the repulsive screened Coulomb parameter, which can have a value between 0.1 and 0.15 but for intermetallic superconductors a value of 0.13 is typically used. Using $T_{\mathrm{c}}$ and $\Theta_{\mathrm{D}}$ taken from Fig.~\ref{HC}(b), a value of $\lambda_{\mathrm{el-ph}} = 0.67 \pm 0.02$ is obtained, suggesting this a moderately coupled superconductor. Using this value of $\lambda_{\mathrm{el-ph}}$ and Eq.~(\ref{mband}) a value for $m^{*}_{\mathrm{band}}$ can be found, as seen in Table~\ref{Elecprop}. Recently, these parameters have also been determined for the related compound Re$_{6}$Hf \cite{Re6Hfa,Re6Hfb}. By substituting Zr by Hf the spin-orbit coupling should be enhanced, and it was hoped that this would provide an increase in the contribution of the spin-triplet component in the superconducting ground state. From the measurements performed in Refs.~\onlinecite{Re6Hfa} and~\onlinecite{Re6Hfb} it is clear that Re$_{6}$Hf and Re$_{6}$Zr are very similar and that the spin-orbit-coupling strength seems to have little effect on the properties of polycrystalline samples at least.
Uemura~\textit{et al}. have described a method for classifying superconductors based on the ratio of the critical temperature $T_{\mathrm{c}}$ to the effective Fermi temperature $T_{\mathrm{F}}$ \cite{Uemera3}. The values of $m^{*}$ and $n$ taken from Table~\ref{Elecprop} can used to calculate an effective Fermi temperature for Re$_6$Zr using
\begin{equation}
k_{\mathrm{B}} T_{\mathrm{F}} = \dfrac{\hbar^{2}}{2 m^{*}} \left( 3 \pi^{2} n \right)^{2/3},
\end{equation}
and the result is presented in Table~\ref{Elecprop}. It has been observed that the high-$T_{\mathrm{c}}$, organic, heavy-fermion, and other unconventional superconductors lie in the range $0.01 \leq T_{\mathrm{c}}/T_{\mathrm{F}} \leq 0.1$ \cite{Uemera1,Uemera2,Uemera3}. However, Re$_6$Zr lies outside of the range for unconventional superconductivity, supporting the view that the superconducting mechanism is primarily conventional. 

\begin{table}[tb!]
\caption{Normal-state and superconducting properties of Re$_{6}$Zr.}\label{tab:a}
\begin{center}
 \begin{tabular}{l l r} 
 \hline\hline
  Re$_{6}$Zr property & Units & Value \\ 
 \hline
 & & \\
 $T_{\mathrm{c}}$ & K & $6.75 \pm 0.05$ \\ 

 $\rho_{0}$ & $\mu\Omega$~cm & $142 \pm 2$  \\

 $\rho_{sat}$ & $\mu\Omega$~cm & $167 \pm 1$ \\
 
 $\Theta_{\mathrm{R}}$ (from resistivity) & K & $237 \pm 2$  \\
 
 $\Theta_{\mathrm{D}}$ (from Sommerfeld coefficient) & K & $338 \pm 9$  \\
 
 $\Theta_{\mathrm{D}}$ (from Debye-Einstein fit) & K & $258 \pm 1$  \\ 
 
 $T_{\mathrm{E}}$ & K & $652 \pm 12$  \\   
 
 $\gamma_{\mathrm{n}}$ & mJ mol$^{-1}$K$^{-2}$ & $26.9 \pm 0.1$ \\ 
 
 $\beta_{3}$ & mJ mol$^{-1}$K$^{-4}$ & $0.35 \pm 0.02$ \\
 
 $\beta_{5}$ & $\mu$J mol$^{-1}$K$^{-6}$ & $1.6 \pm 0.1$ \\
 
 $\lambda_{\mathrm{el-ph}}$ & & $0.67 \pm 0.02$ \\
 
 $\Delta C/\gamma_{\mathrm{n}} T_{\mathrm{c}}$& & $1.60 \pm 0.02$ \\
 
 $\Delta_{0}/k_{\mathrm{B}}T_{\mathrm{c}}$& & $1.86 \pm 0.05$ \\

 $\mu_0H_{\mathrm{c1}}\left(0\right)$ & mT & $10.3 \pm 0.1$ \\ 
 
 $\mu_0H_{\mathrm{c2}}\left(0\right)$ & T & $11.2 \pm 0.2$ \\
 
 $\mu_0H^{\mathrm{cal}}_{\mathrm{c}}\left(0\right)$ & mT & $175 \pm 3$  \\ 
 
  $\mu_0H^{\mathrm{exp}}_{\mathrm{c}}\left(0\right)$ & mT & $130 \pm 2$  \\ 
 
 $\mu_0H_{\mathrm{c}2}^{\mathrm{orbital}}\left(0\right)$ & T & $11.41 \pm 0.05 $ \\
 
 $\mu_0H_{\mathrm{c}2}^{\mathrm{Pauli}}\left(0\right)$ & T & $12.35 \pm 0.09$  \\ 

 $\xi_{\mathrm{GL}}\left(0\right)$ & nm & $5.37 \pm 0.09$ \\
 
 $\lambda_{\mathrm{GL}}\left(0\right)$ & nm & $247 \pm 4$ \\
 
 $\kappa_{\mathrm{GL}}\left(0\right)$ &  & $46.2 \pm 0.8$ \\

 & & \\
 \hline\hline
\end{tabular}
\end{center}
\end{table}

\section{Summary}

In summary, single-phase polycrystalline samples of Re$_{6}$Zr were prepared by the arc-melting technique. Powder x-ray diffraction data confirmed the cubic, noncentrosymmetric $\alpha$-Mn crystal structure and the phase purity of the samples. The normal-state and superconducting properties of Re$_{6}$Zr were studied using magnetization, heat-capacity, and resistivity measurements. We have established that Re$_6$Zr is a moderately coupled superconductor with a transition at ${T_{\mathrm{c}}=\left(6.75\pm 0.05\right)~\mathrm{K}}$. In the normal state, resistivity measurements show that Re$_{6}$Zr has poor metallic behavior that is dominated by disorder. We showed that it is possible to fit these data with a parallel-resistor model that considers contributions in addition to the electron-phonon interactions. Specific-heat measurements of the normal state reveal no indication of any structural phase transitions down to low temperature and were fit using a simple Debye-Einstein function. The jump in specific heat at $T_{\mathrm{c}}$ is ${\Delta C/\gamma_{\mathrm{n}} T_{\mathrm{c}} = 1.60 \pm 0.02}$, while $C\left(T\right)$ below $T_{\mathrm{c}}$ was fit using the BCS model, giving $\Delta_{0}/k_{\mathrm{B}}T_{\mathrm{c}}=1.86\pm0.05$. Both values are well above those expected for a conventional BCS superconductor, suggesting the electron-phonon coupling is enhanced in this system. The mean free path $\ell_{\mathrm{e}}$ is estimated to be $\left(1.45\pm0.02\right)$~nm. The best approximation for $H_{\mathrm{c}2}\left(0\right)$ was found using the WHH model. From $H_{\mathrm{c}2}\left(0\right)$ the coherence length was calculated with ${\xi_{\mathrm{GL}}\left(0\right) = \left(5.37\pm0.09\right)~\mathrm{nm}}$, confirming that Re$_{6}$Zr is in the dirty limit. Using the magnetization data, it was possible to estimate $\mu_{0}H_{\mathrm{c1}}\left(0\right) = \left(10.3\pm 0.1\right)$~mT and so calculate the penetration depth $\lambda_{\mathrm{GL}}\left(0\right) = \left(247\pm4\right)$~nm. The Ginzburg-Landau coefficient $\kappa_{\mathrm{GL}} \left(0\right)= 46.2\pm 0.8$ confirmed that Re$_{6}$Zr is a strong type-II superconductor. A summary of all the experimentally measured and estimated parameters is given in Table~\ref{tab:a}. From our measurements we can conclude the superconducting order parameter is well described by an isotropic gap with $s$-wave pairing symmetry and enhanced electron-phonon coupling, despite the observation of spontaneous magnetization associated with TRS breaking being observed at temperatures below the superconducting transition in previous work~\cite{Re6Zr}. This suggests Re$_{6}$Zr has a superconducting ground state that features a dominant $s$-wave component, while the exact nature of the triplet component is undetermined. In order to determine if the superconductivity is nonunitary, further experimental work on high-quality single crystals, as well as further analysis of ``clean'' and ``dirty'' samples to examine the role grain boundaries and impurities play in determining the superconducting behavior of Re$_{6}$Zr, is vital.

\begin{acknowledgments}
This work is funded by the EPSRC, United Kingdom, through Grants No. EP/I007210/1 and No. EP/M028771/1. 
\end{acknowledgments}

\bibliography{Re6Zr_DM_References}

\end{document}